\documentclass[useAMS,usenatbib]{mn2e}
\usepackage{amssymb,epsfig}
\usepackage{afterpage}
\newcommand{\bc}{\begin{center}}
\newcommand{\ec}{\end{center}}
\newcommand{\be}{\begin{equation}}
\newcommand{\ee}{\end{equation}}
\newcommand{\ba}{\begin{array}}
\newcommand{\ea}{\end{array}}
\newcommand{\bea}{\begin{eqnarray}}
\newcommand{\eea}{\end{eqnarray}}
\def\ga{\mathrel{\mathchoice {\vcenter{\offinterlineskip\halign{\hfil
$\displaystyle##$\hfil\cr>\cr\sim\cr}}}
{\vcenter{\offinterlineskip\halign{\hfil$\textstyle##$\hfil\cr
>\cr\sim\cr}}}
{\vcenter{\offinterlineskip\halign{\hfil$\scriptstyle##$\hfil\cr
>\cr\sim\cr}}}
{\vcenter{\offinterlineskip\halign{\hfil$\scriptscriptstyle##$\hfil\cr
>\cr\sim\cr}}}}}
\def\la{\mathrel{\mathchoice {\vcenter{\offinterlineskip\halign{\hfil
$\displaystyle##$\hfil\cr<\cr\sim\cr}}}
{\vcenter{\offinterlineskip\halign{\hfil$\textstyle##$\hfil\cr
<\cr\sim\cr}}}
{\vcenter{\offinterlineskip\halign{\hfil$\scriptstyle##$\hfil\cr
<\cr\sim\cr}}}
{\vcenter{\offinterlineskip\halign{\hfil$\scriptscriptstyle##$\hfil\cr
<\cr\sim\cr}}}}}
\def\mkm{{\mu}\rm{m}}

\def\degr{\hbox{$^\circ$}}

\title[Interstellar extinction and polarization]
{Interstellar extinction and polarization ---  
A spheroidal dust grain approach perspective}
\author[Das et al.]{H.K. Das$^{1}$\thanks{E-mail:
hkdas@iucaa.ernet.in (HKD); nvv@astro.spbu.ru (NVV); ilin55@yandex.ru (VBI)},
N.V. Voshchinnikov$^{2}$ and V.B. Il'in$^{2,3}$\\
$^{1}$IUCAA, Post Bag 4, Ganeshkhind, Pune 411 007, India\\
$^{2}$Sobolev Astronomical Institute, St. Petersburg University, St. Petersburg 198504, Russia \\
$^{3}$Main (Pulkovo) Astronomical Observatory, St. Petersburg 196140, Russia}

\begin{document}

%\date{in original form 2009 ...}
\date{Received 2009 ...}

\pagerange{\pageref{firstpage}--\pageref{lastpage}} \pubyear{2009}

\maketitle

\label{firstpage}

\begin{abstract}

{
 We extend and investigate the spheroidal model of interstellar dust grains
used to simultaneously interpret the observed 
interstellar extinction and polarization curves.
% We consider imperfectly aligned spheroids with a simple power-law size distribution.
 We compare our model with similar models recently suggested by other authors,
study its properties and apply it to fit 
the normalized extinction $A(\lambda)/A_{\rm V}$ and the 
polarizing efficiency $P(\lambda)/A(\lambda)$ measured in
the near IR to far UV region for several stars seen through one large cloud.
 We conclude that the model parameter $\Omega$ being the angle between the
line of sight and the magnetic field direction
can be more or less reliably determined from comparison
of the theory and observations.
 This opens a way to study the spatial structure of
interstellar magnetic fields by using
multi-wavelength photometric and polarimetric observations.
}
\end{abstract}


\section{Introduction}

Modelling of the wavelength dependencies of interstellar
extinction $A(\lambda)$ and polarization $P(\lambda)$
along with analysis of infrared bands is
the main source of information about the properties of interstellar
grains (\citealt{draine_2003}; \citealt{whitt_2003}; \citealt{vosh_2004}).

Extinction data were often used to constrain the size distribution
of carbonaceous and silicate bare grains (e.g. \citealt{mathis_1977};
\citealt{zubko_1996}; \citealt{weind_2001}).
 These data were also modeled with
composite particles %consisting of several materials
(\citealt{vaidya1}; \citealt{vosh_2006}; Iat\`i \citealt{iati_2008}).
 In all these studies the grains were assumed to be spherical,
i.e. their shape was not involved.

To get information about the grain shapes as well as
the ambient magnetic fields
one needs to consider interstellar polarization.
 Interpretation of its wavelength dependence
includes calculations of the extinction cross sections of
rotating partially aligned non-spherical particles.
 In early studies (e.g. \citealt{hong_1980}; \citealt{vosh_1989})
an unphysical model of infinitely long
cylinders was applied\footnote{ 
  \citet{lg97} also considered finite cylinders
  for modelling interstellar extinction and polarization.}.
 More realistic is the spheroidal model of grains with %where
the shape of these axisymmetric particles being %is 
characterized by the only parameter ---
the ratio of the major to minor semi-axis $a/b$.
 The optical properties of spheroids have long been of
interest ({\rm e.g., \citealt{martin_1978}; \citealt{rogers_1979};
\citealt{onaka_1980}; \citealt{draine_1984}; \citealt{vosh_1993};
\citealt{kim_1995})}.
 The growth of such grains in the {\rm interstellar medium (ISM)}
has been discussed by \citet{stark_2006},
their absorption and scattering properties (in the Rayleigh limit)
by \citet{min_2006} and mineralogy
by \citet{li_2007}.

Over the past decade spheroids have found wide applications
in studies of optics of
cometary dust ---
\citet{das_2006}, \citet{moreno_2007},
circumstellar dust --- \citet{wolf_2002},
interplanetary dust --- \citet{lasue_2007}.
 \citet{greenberg_1996} used prolate spheroids and \citet{lee_1985},
\citet{hild_1995} oblate ones to interpret the silicate
and ice polarization features of the BN object.
 \citet{draine_2006} predicted
halos due to X-ray scattering by oblate particles,
detection of which would serve as a test for dust models.

{\rm  The spheroidal model of cosmic dust grains is particularly promising}
for interpretation of the interstellar polarization and extinction data.
 A detailed review of early works on the subject can be found in \citet{vosh_2004},
so we consider recent applications of the model.
\citet{gupta_2005}
well reproduced the average Galactic
extinction curve in the region $0.3 - 9.5\ \mkm^{-1}$ with
silicate and graphite oblate spheroidal ($a/b=1.33$) particles
rotating in a plane and having the size distribution
$n(r_V) \propto r_V^{-3.5}$ with $r_V = 0.005 - 0.25\ \mkm$
where $r_V$ is the radius of a volume equivalent sphere.
 They found that the shape of grains did not affect the
abundance problem, but essentially changed the polarization
{\rm efficiency factors} at different particle orientation angles.
 It should be noted
that interpretation of the extinction data alone is usually made within
the spherical model of dust grains (see for details the recent
{\rm review of \citealt{draine_2009})}.
\citet{draine_2006}
{\rm and}
\citet{drainef_2009}
fitted both
the average Galactic extinction and polarization curves
in the range $0.4 - 9.5\ \mkm^{-1}$
applying picket-fence oriented silicate and graphite oblate spheroids
{\rm with the aspect ratio in the interval $a/b=1-2$
with}
specific size distributions.
 The grain alignment degree was size dependent and 
showed a sharp increase at $r_V \sim 0.1\ \mkm$.
 Voshchinnikov \& Das~(2008; hereafter VD08)
studied the wavelength dependence of the
ratio of the linear polarization degree to extinction %$P(\lambda)/A(\lambda)$
{\rm (so called polarizing efficiency)}
from the ultraviolet to near-infrared.
 They found that the wavelength dependence of $P(\lambda)/A(\lambda)$
was mainly determined by the particle composition and size whereas
the absolute values of this ratio %$P(\lambda)/A(\lambda)$ 
depended on the {\rm particle} shape, degree and direction of alignment.

In this paper we investigate further advantages of
the spheroidal approach to interstellar grains.
 In contrast to other works we calculate the optical
properties of spheroids of all sizes exactly even
{\rm at  far UV wavelengths,
consider ensembles of particles with various degree and direction of alignment,}
compare different approaches to extinction and polarization curve modelling,
{\rm treat}
the observational data in a new way 
(the wavelength dependence of the polarizing efficiency is fitted instead 
of the polarization curve),
apply the theory to observational data for concrete stars, and
discuss a possibility to determine the direction
of interstellar magnetic fields from the data.
 In Section 2 we describe details of our model and 
compare it with the models used earlier. 
 In Section 3 the results obtained from our model are compared
with observations of seven stars 
for which the line of sight intersects one main cloud,
keeping in view %the fact 
that ``mean extinction curves as well as mean
total-to-selective extinction ratios
are possibly averaged over several interstellar clouds of possibly different
physical parameters'' \citep{krelo_1989}.
 Section 4 contains our conclusions.

%%%%%%%%%%%%%%%%%%
\section{Theory}\label{theor}

Our modelling is based on the simple standard
assumptions concerning dust grain materials, size distribution, 
shape and alignment.
 Homogeneous spheroids considered are characterized by their type
(prolate or oblate), the aspect ratio $a/b$ and a size parameter.
 As the latter we choose
the radius $r_V$ of a sphere whose volume is equal to that of
a non-spherical particle, i.e.
$r_V^3  = a b^2$ for prolate spheroids and
 $r_V^3  = a^2 b$ for oblate ones.

\subsection{Dust grain materials}\label{materials}

Interstellar dust mainly consists of
five heavy
elements: C, O, Mg, Si and Fe, which
are locked in carbonaceous and silicate particles
(\citealt{jones99}; \citealt{draine_2009}).
 We consider sub-micron grains from amorphous carbon (AC1) and
amorphous silicate (astronomical silicate, astrosil).
 In order to reproduce the 2175\AA\, absorption feature
 small graphite spheres
 { with radius $r_{\rm gra}=0.02\,\mu$m}
 are also involved\footnote{ 
    Note that graphite is not the only material considered as the carrier of the 
    the 2175\AA\, absorption feature (see, e.g., \citealt{Li_2008})}.
 The optical constants of the materials are taken from
\citet{rouleau_1991} and \citet{laordr93}.

\subsection{Size distribution}\label{s-z}

Interpretation of the interstellar extinction is usually aimed
at reconstruction of the dust grain size distribution
consistent with
the cosmic abundances and some other constraints
(see discussion in \citealt{vosh_2004}).

We choose a power-law size distribution %function
\be
n(r_V) \propto r_V^{-q}\,,
\label{szd}
\ee
which was derived by \citet{mathis_1977} (see also \citealt{draine_1984})
from minimization of $\chi^2$-statistic.
 The distribution has three parameters:
the lower ($r_{V, \min}$) and upper ($r_{V, \max}$) limits
and the power index $q$.
 It was widely used in modelling of interstellar extinction and
radiative transfer in dusty objects.
 The average Galactic extinction curve can be reproduced
with the so-called standard MRN mixture ---
an ensemble of carbonaceous (graphite) and silicate spheres
(in nearly equal proportions)
with the parameters: $q=3.5$, $r_{V, \min}\approx 0.005 \,\mu$m
and $r_{V, \max} \approx 0.25 \,\mu$m.

\subsection{Cross section averaging}

\begin{figure}
\centering
   \includegraphics[width=8cm]{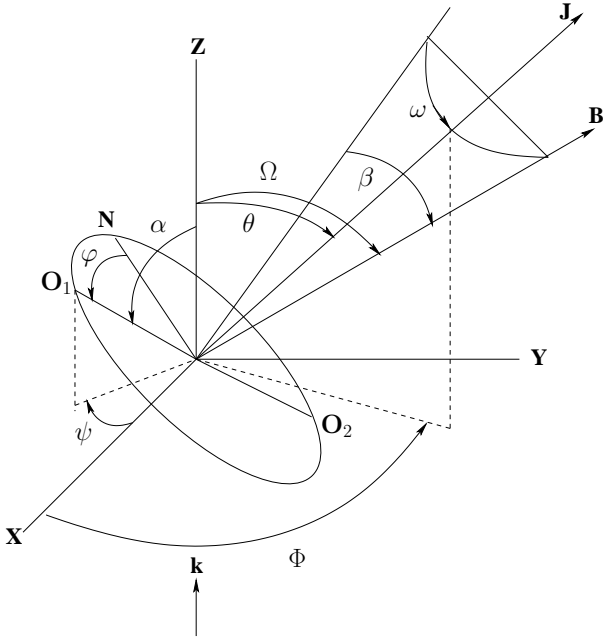}
 \caption{Geometrical configuration of a spinning and
 wobbling prolate spheroidal grain.
  The major (symmetry) axis of the particle %a spheroid axis
 O$_1$O$_2$ is situated in the spinning plane NO$_1$O$_2$
 which is perpendicular to the the angular momentum {\rm J}.
 The direction of light propagation {\rm k} is parallel to the $Z$-axis
 and makes the angle $\alpha$ with the particle symmetry axis.
 %? The sky is represented by the $XY$ plane.
 }
   \label{f-sky}
\end{figure}

Let us consider non-polarized light passing through a cloud
that contains three populations of grains:
rotating spheroids from amorphous silicate {\rm (Si),}
amorphous carbon (C) 
and graphite (gra) spheres.
 For a line of sight, extinction (in stellar magnitudes) and
linear polarization (in percentage) produced by the cloud 
can be written
as
\begin{equation}
A({\lambda})= 1.086 N_{\rm d} \langle C_{\rm ext} \rangle _{\lambda}\,,
\,\,\,\,\,\,\,\,\,
P({\lambda})=N_{\rm d} \langle C_{\rm pol} \rangle_{\lambda}100{\%}\,.
\label{eq2}
\end{equation}
 Here,
$N_{\rm d}=N_{\rm C}+N_{\rm Si}+N_{\rm gra}$
is the total dust grain column density,
$ \langle C_{\rm ext} \rangle_{\lambda}$ and
$ \langle C_{\rm pol} \rangle_{\lambda}$ are
the extinction and polarization cross sections, respectively,
averaged over the grain populations, i.e.
{
\be
 \langle C_{\rm ext} \rangle_{\lambda}  =  {\cal K}_{\rm C} \overline{C}_{\rm ext, C}({\lambda})
  +
 {\cal K}_{\rm Si} \overline{C}_{\rm ext, Si}({\lambda}) +
 {\cal K}_{\rm gra} {C}_{\rm ext, gra}({\lambda})\,,
\label{caver}
\ee
\be
 \langle C_{\rm pol} \rangle_{\lambda}  =  {\cal K}_{\rm C} \overline{C}_{\rm pol, C}({\lambda})
 +  {\cal K}_{\rm Si} \overline{C}_{\rm pol, Si}({\lambda})\,,
\label{cpaver}
\ee
%where ${\cal K}_{i}=N_i/N_{\rm d}$ ($i \equiv {\rm C, Si, gra})$ is
where
$
{C}_{\rm ext, gra}({\lambda}) = \pi r^2_{\rm gra} {Q}_{\rm ext, gra}({\lambda})
$
and
${\cal K}_{i}=N_i/N_{\rm d}$ ($i \equiv {\rm C, Si})$ is
the relative column density of the $i$-th population.
 It is evident that
$$
{\cal K}_{\rm C}+{\cal K}_{\rm Si}+{\cal K}_{\rm gra}=1.
$$
 The values of $\overline{C}_{{\rm ext, pol}, i}$ are obtained
by averaging of the cross sections
over the size distribution and grain orientations
{\rm (\citealt{hong_1980}; \citealt{vosh_1989})}
\bea
 \overline{C}_{{\rm ext}, i}({\lambda}) & = & { \left(\frac{2}{\pi} \right)^2}
{\int_{r_{V, \min, i}}^{r_{V, \max, i}}}
{\int_{0}^{\pi/2}}{\int_{0}^{\pi/2}}{\int_{0}^{\pi/2}} \pi r_{V, i}^2 \,Q_{{\rm ext}, i}({\lambda}) \,
 \nonumber \\ & \times &
n_i(r_V) \, f(\xi, \beta) \,  d{\varphi} d{\omega} d{\beta} d{r_V}\,,
\label{eq3}
\eea
\bea
 \overline{C}_{{\rm pol}, i}({\lambda}) & = & {\frac{2}{\pi^2}}
{\int_{r_{V, \min, i}}^{r_{V, \max, i}}}
{\int_{0}^{\pi/2}}{\int_{0}^{\pi}}{\int_{0}^{\pi/2}}\pi r_{V, i}^2 \, Q_{{\rm pol}, i}({\lambda})\,
 \nonumber \\ & \times &
n_i(r_V) \, f(\xi, \beta) \, \cos 2{\psi} \, d{\varphi} d{\omega} d{\beta} d{r_V} \,.
\label{eq4}
\eea
}
Here, ${\beta}$ is the precession-cone angle for
the angular momentum {\rm J} which spins around
the direction of the magnetic field {\rm B},
${\varphi}$ the spin angle,
${\omega}$  the precession angle (see Fig.~\ref{f-sky}).
%%%%%%%%%%%%%%%%%%%%%%%%%%%%%%%%%%%%%%%%%%%%
%%%%%%%%%%%%%%%%%%%%%%%%%%%%%%%%%%%%%%%%%%%%
 The quantities $Q_{\rm ext}=(Q_{\rm ext}^{\rm TM} + Q_{\rm ext}^{\rm TE})/2$
and  $Q_{\rm pol}=(Q_{\rm ext}^{\rm TM} - Q_{\rm ext}^{\rm TE})/2$ are
the extinction and polarization efficiency factors for
non-polarized incident radiation (\citealt{vosh_1993}),
\be
\cos 2\psi = \left\{
\begin{array}{l@{\quad {\rm ,} \quad} l}
\cos 2\Phi \left[\frac{2\cos^2\varphi \cos^2\theta}
{1-\cos^2\varphi \sin^2\theta} -1 \right] & {\rm prolate} \\
\cos 2\Phi & {\rm oblate,}
\end{array} \right.
\ee
and
\be
\cos\theta = \cos\Omega \cos\beta + \sin\Omega \sin\beta \cos\omega,
\ee
\be
\cos 2\Phi = \left[ \frac{2\sin^2\beta \sin^2\omega}{\sin^2\theta} - 1 \right],
\ee
\be
\cos\alpha = \left\{
\begin{array}{l@ {\quad {\rm ,}  \quad} l}
\sin\theta \cos\varphi & {\rm prolate}\\
\sqrt{1-\sin^2\theta \cos^2\varphi} & {\rm oblate.}
\end{array} \right. 
\ee
Here, $\Omega$ is the angle between the line of sight
and the magnetic field direction ($0\degr \leq \Omega \leq 90\degr$).
 The value $\Omega=90^\circ$ corresponds to the
case when the particle rotation plane contains
the light propagation vector {\rm k},
which gives the maximum degree of linear polarization.
 For $\Omega=0^\circ$,
the light falls perpendicular to the particle rotation plane
and from symmetry reasons
the net degree of polarization produced is zero.

The efficiency factors were calculated by using
a solution to the light scattering problem for a homogeneous spheroid
obtained by the method of separation of
variables in spheroidal coordinates \citep{vosh_1993}
and a modern approach to computations
of spheroidal wave functions \citep{vosh_2003}.
 In contrast to all works  
where the spheroidal model was recently applied 
(\citealt{gupta_2005}; \citealt{draine_2009}; {\rm \citealt{drainef_2009}})
we computed the optical properties of the particles of all sizes exactly
even in the far UV {\rm spectral} region.

\subsection{Grain alignment}\label{alignment}

 Interstellar non-spherical grains should be partially aligned.
 The ``picket fence'' or perfect rotating alignment
often used in modelling {\rm of polarization}
(e.g. \citealt{kim_1995}; {\rm \citealt{drainef_2009}}) is
a crude approximation giving a too high polarizing efficiency
in comparison {with} the empirical upper limit
{\rm (e.g., \citealt{greenberg_1978})}
$$
P_{\max}/A_{\rm V} \la 3\%/{\rm mag.}
$$

 Although the theory of grain alignment is actively discussed
(\citealt{laz_2007}) it still has little practical significance.
 Therefore, we consider the classical alignment mechanism~---
so-called imperfect Davis--Greenstein (IDG) orientation (\citealt{davis_1951}).
 The IDG mechanism is described by the function ${f}(\xi, \beta)$
depending on the alignment parameter $\xi$ and the precession angle $\beta$
{\rm (\citealt{hong_1980}; \citealt{vosh_1989})}
\begin{equation}
{f}(\xi, \beta) = \frac{\xi \sin \beta}{(\xi^2 \cos^2 \beta  + \sin^2 \beta)^{3/2}}.
\label{eq5} 
\end{equation}
 The parameter $\xi$ depends on the particle size $r_V$,
the imaginary part of the grain magnetic susceptibility
%\]
$\chi'' ={\varkappa}{\omega_{\rm d}} /T_{\rm d}$\,,
%\]
where $\omega_{\rm d}$ is the angular velocity of a grain
and $\varkappa$ = $2.5 \times 10^{-12}$ (\citealt{davis_1951}),
gas density $n_{\rm H}$,
the strength of magnetic field $B$,
and temperatures of dust $T_{\rm d}$ and gas $T_{\rm g}$.
 By introducing a parameter %$\delta_0$
\begin{equation}
\delta_0 = 8.23 \times 10^{23} \frac{{\varkappa B^2}}
{n_{\rm H}T_{\rm g}^{1/2}{T_{\rm d}}},~\mu{\rm m},
\label{eq7} 
\end{equation}
one can get
\begin{equation}
\xi^2  = \frac{r_V + \delta_0 (T_{\rm d}/T_{\rm g})}{r_V +\delta_0} .
\label{eq6}
\end{equation}

We assume that carbon and silicate grains have similar
alignment but different size distributions.
 Thus, our model has 10 main parameters:
$r_{V, \min}$, $r_{V, \max}$ and $q$ for carbon and silicate
grains,  relative  density for carbon
(${\cal K}_{\rm C}$) and silicate (${\cal K}_{\rm Si}$) grains,
degree ($\delta_0$) and direction ($\Omega$) of grain alignment.

\subsection{Comparison with other models}\label{oth-m}

There were just a few approaches to simultaneous modelling of the
wavelength dependencies of interstellar extinction and linear polarization.
 In most papers where an
{\rm imperfect
 alignment of grains
was involved (along with a power-law size distribution) 
the authors derived
that}
particles with $r_V \la 0.1 \mkm$ should be poorly (randomly) oriented
while those with $r_V \ga 0.1 \mkm$ should be nearly perfectly aligned 
(e.g., \citealt{kim_1995}; \citealt{drainef_2009}).
 Some physical grounds for such alignment of interstellar grains come, e.g., from 
\citet{mathis_1986}, \citet{goodman_1995}, \citet{martin_1995}.

{
As the actual composition of carbonaceous grains in the ISM 
is still unknown  (\citealt{drainef_2009}),} 
%and about 85\% of carbon may be in aromatic structures 
%(\citealt{pendleton_2002}), 
in interstellar dust modelling
one often considered large (sub-micron) graphite grains instead of 
amorphous carbon ones.
 As polarization produced by such graphite spheroids
may be
{\rm to a certain extent}
peculiar, 
the carbonaceous grains were often assumed to be either
spherical or randomly aligned 
(e.g., \citealt{mathis_1986};  \citealt{drainef_2009}).

{ 
To compare these approaches with our one, we have made some calculations
for {\em three additional models}:
%\begin{itemize}

%\item %[model A1]
model A1~-- 
imperfectly DG-aligned silicate and {\em graphite} spheroids;
%\\

%\item %[model A2]
model A2~-- 
imperfectly DG-aligned silicate spheroids and 
   {\em randomly oriented} amorphous carbon spheroids; 
%   \\

%\item %[model A3]
model A3~-- 
silicate and amorphous carbon spheroids with
   {\em randomly oriented small particle} and %with $r_V < r_{V,cut}$ and
   perfectly DG-aligned large particle. % with $r_V \geq r_{V,cut}$.
%\end{itemize}   

The grain size distributions and the population of small graphite particles 
are not varied, i.e. they are the same as in our basic model described in 
Sects.~\ref{materials}--\ref{alignment}. }

 When considering graphite spheroids in the A1 model, we use
the standard {\rm 2/3--1/3} approximation which gives reasonably accurate
results (\citealt{draine_1993}). 
 The orientation function  for amorphous carbon grains in the A2 model 
is ${f}(\beta) = \sin\beta$. 
In the A3 model this function is size dependent
\begin{equation}
{f}(\beta,r_V) = \left\{
\begin{array}{ll}
 \sin \beta & {\rm for}\ \ r_V \le r_{V,\rm cut} \\
 \delta(\beta) & {\rm for}\ \ r_V > r_{V,\rm cut}
\end{array} , \right.
\label{eq6a}
\end{equation}
where $\delta(\beta)$ is the Dirac delta function,
$r_{V,\rm cut}$ a cut-off parameter.
 
So, we assume that all particles whose volume is
smaller than that of a sphere with the radius $r_{V,\rm cut}$ 
are randomly oriented, while other particles have
perfect rotation alignment, i.e. their angular moment
vectors are parallel to the magnetic field direction. 
In our calculations we choose $r_{V,\rm cut}$ to be 0.11 $\mkm$
(cf. \citealt{drainef_2009}).
    
%%%%%%%%%%%%%%%%%%%%%%%%%%%%%%%%%%%%%%%%%%%%
\begin{figure*}

\includegraphics[angle=0, width=15cm]{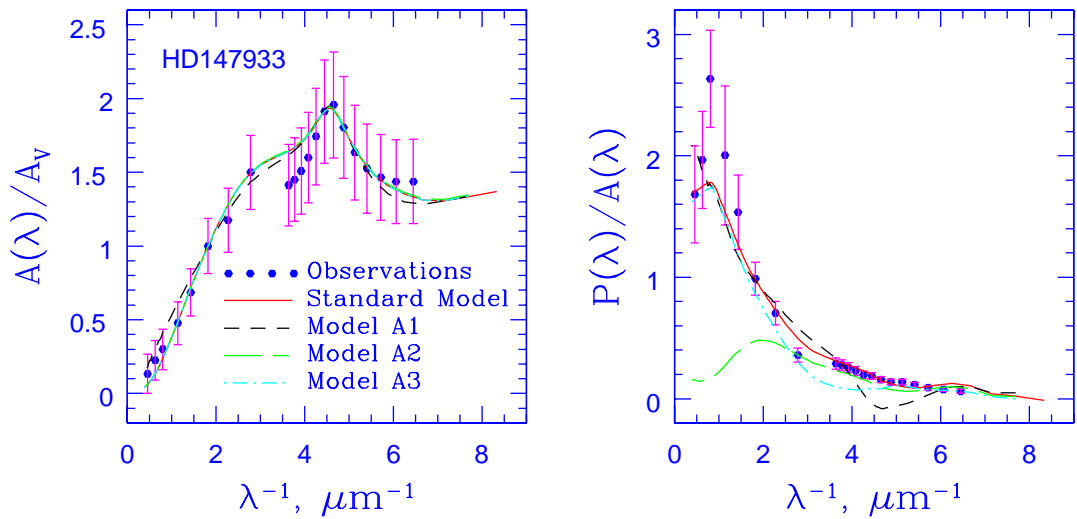}
 \caption{Normalized extinction (left panel) and polarizing efficiency (right panel)
for our basic model and three additional models 
%considered in the text.
with graphite spheroids (A1);
randomly oriented amorphous carbon spheroids (A2);
small randomly oriented silicate and amorphous carbon spheroids (A3)
 (see the text for mode details).
 Oblate spheroids with $a/b = 3$ and $\Omega = 33^\circ$,
for other parameters see the last line in Table~\ref{t933}.
 Observational data for HD 147933. 
}
\label{otherm}
\end{figure*}
%%%%%%%%%%%%%%%%%%%%%%%%%%%%%%%%%%%%%%%%%%%%

 Some results obtained with the models considered
are presented in Fig.~\ref{otherm},
the fitting procedure used is described in the next Section. 
 We see that all models can give very similar
extinction curves that well fit the observational data.
 Only the A1 model with large graphite particles provides 
the curve that differs a bit from others.
 Certainly,
 the extinction data are not enough to constrain
the models.

{\rm
 The model dependencies of  polarizing efficiency
shown in the right panel of the figure
vary in form.
 In particular, the model with randomly oriented
carbonaceous particles produces too small polarization in the visual
and IR parts of spectrum.
Apparently, in this case  a better fit can be reached if one
takes a more complex size distribution function or (and)
increases the alignment efficiency
of silicate grains.
However, this complicates the model, therefore, we prefer
our  model having a small number
of the parameters and simple assumptions.
}

%@@@@@@@@@@@@@@@@@@@@@@@@@@@@@@@@@@@@@@@@@@@@@@@@@@@@@@@@@@@@@@@@@@@

%\afterpage{\clearpage}
%%%%%%%%%%%%%%%%%%%%%%%%%%%%%%%%%%%%%%%%%%%%%%%%%%%%%%%%%%%%%%%%%%%%%%%%
\section{Comparison with observational data}
%%%%%%%%%%%%%%%%%%%%%%%%%%%%%%%%%%%%%%%%%%%%%%%%%%%%%%%%%%%%%%%%%%%%%%%%

So far, most efforts in modelling of the optics of interstellar dust
were directed at fitting of 
the average Galactic extinction curve (often with some additional constraints)
by making use of spherical particles.
 Interpretation of the curves observed towards individual objects
was very seldom, and that of the extinction and polarization curves 
{\rm was done  only for a few objects.
\citet{aei87} applied the  model of
infinitely long coated cylinders with the imperfect DG-alignment
to the interpretation
of the wavelength dependence of interstellar extinction, linear and
circular polarization for the star HD~204827 in the visual and IR parts
of spectrum.
\citet{kim_1994}
used infinitely long cylinders with the perfect DG-alignment
($\Omega = 90^\circ$) to interpret the data for HD 25443 and
HD 197770.
 Infinitely long cylinders with  a silicate core and organic refractory
mantle were chosen by \citet{lg98}
for the interpretation of the extinction
and polarization curves for HD~210121.
 So, \citet{vosh_2008} {were first who} utilized  the spheroidal model
{to consider} both the extinction and polarization wavelength dependencies 
observed for two stars.}
 
\subsection{Sample of objects}\label{sel-o}

For application of our model we
{\rm chose}
stars seen through a single cloud
with the extinction and polarization curves known 
in a wide wavelength region.
 Initially,
 {\rm preference was given to}
 five stars (HD~24263, HD~62542,
HD~99264,  HD~147933 and  HD~197770) 
with measured UV polarization \citep{ander_1996}.
 These stars are not peculiar and not very distant.
 Later, two stars (HD~147165 and HD~179406) for which
the polarization data in the visible were obtained by 
us was added to our sample.

 In this paper we discuss
 {\rm  extinction and polarization of
 six stars
(HD~24263, HD~99264, HD~147165, HD~147933, 
HD~179406, and  HD~197770).
 Two objects (HD~62542 and HD~147933) 
were considered in our previous paper (VD08).}
 Almost all the stars are not very far away from 
the galactic plane {\rm (Table~\ref{t_st1})}
and are in dusty environment.   
 One can assume 
that these stars are seen through single dust clouds 
as they have weak rotation angle
of linear polarization \citep{ander_1996}
and narrow symmetric sodium D$_1$ and D$_2$ lines \citep{zubko_1996}.
 A few comments on the stars should be made.

\smallskip
\noindent{\it HD~24263} {
is a double star (the secondary is a B9 star)
seen through a large cloud AG6 located between the Local Bubble  
and Orion--Eridanus Super-bubble at distance $\sim$150 pc and
observed in NaI D and CaII lines 
(\citealt{Genova_2003}; \citealt{Welsh_2005}).
One also detects the molecules CH, CH$^+$ and CN in the line of site
(\citealt{Penprase_1990}).
}

\smallskip
\noindent{\it HD~99264}
 is located in the region of the Scorpios--Centaurs OB association
and the Chameleon--Musca dark clouds
(\citealt{Platais_1998}; \citealt{Corradi_2004}).
 Consideration of high-resolution NaI D profiles and
$uvby\beta$ photometry for over 60 nearby B stars allows
\citet{Corradi_2004}
to conclude that interstellar gas and dust
in this region
are distributed in two extended sheet-like structures.
 The nearby feature located at the distance $< 60$ pc
provides the colour excess $E(b-y) \approx 0.05$ while
the second feature at 120--150 pc has $E(b-y) \approx 0.2$.

{
\smallskip
\noindent{\it HD~147165} ($\sigma$ Sco) 
is a beta Cephei type pulsating variable  
and a spectroscopic binary with the secondary being a B1V star
(\citealt{North_2007}).
 It is a member of the Upper Scorpius subgroup within the
Sco OB2 association %(North et al., 2007, MNRAS 380, 1276). 
related with reflection nebula in the Ophiuchus cloud complex 
region (\citealt{Mamajek_2008}).
}

{
\smallskip
\noindent{\it HD~147933} is a component of 
the well-studied binary star $\rho$~Oph AB.
 High-resolution KI, NaI, CaII observations show
very complex profiles of atomic lines in its spectrum (\citealt{Snow_2008}).  
 The star is associated with a strong reflection nebula
and a significant amount of dust (and gas) is assumed to lie behind it.
%This is the star $\rho$~Oph AB related
%to the dark cloud T2171 (clump P21) (\citealt{dob05}).
}

{
\smallskip
\noindent{\it HD~179406} (20 Aql) 
is a variable star behind a translucent cloud 
(\citealt{Hanson_1992}; \citealt{dob05}).
Though the absorption and emission lines %observed in its spectrum 
have  at least 3 components, the dominant one is %well 
seen in most of the atomic and all molecular lines.
%appears to be located behind the
%dark cloud T277 (\citealt{dob05}).
}

\smallskip
\noindent{\it HD~197770} 
is an evolved spectroscopic eclipsing binary star 
with both components being B2III stars \citep{Clayton_1996}.
 The object is close to Cyg OB7 and Cep OB2 and probably
lies on the edge of a star formation region including
molecular clouds L1036 and L1049.
 It is associated with non-stellar IRAS sources (see
for more details \citealt{Gordon_1998}).

{
\smallskip
So, the clouds observed in the selected lines of sight
look to be rather different and hence
their dust grain ensemble parameters may also differ.
}

%%%%%%%%%%%%%%%%%%%%%%%%%%%%%%%%%%%%%%%%%%%%%%%%%%%%%%%%%%%%%%%%%%%%%%%%%%%
\begin{table*}
 \centering
 %\begin{minipage}{150mm}
 \caption{Target stars.}\label{t_st1}
  \begin{tabular}{lcrcccc} \hline
~~~Star & $\it l$ &  $\it b$~~ & $D$ (pc) &  Sp & $R_{\rm V}$ & $A_{\rm V}$ \\
\hline
HD~24263 & 182.1 & -34.9   & 171 & B5V & { 3.44 } & { 0.72 } \\
HD~62542 & 255.9 & ~--9.2  & 396 & B5V & 2.74 & 0.99 \\
HD~99264 & 296.3 & --10.5  & 266 & B2.5V & 3.15 & 0.85  \\
HD~147165& 351.3 & +17.0   & 137 & B1III & 3.60 & 1.48  \\
HD~147933& 353.7 & +17.7   & 118 & B2.5V & 4.41 & 2.07   \\
HD~179406& ~28.2 & ~--8.3  & 227 & B3V & 2.88 & 0.89  \\
HD~197770& ~93.9 & ~+9.0   & 943 & B2V & 2.77 & 1.61  \\
\hline
\end{tabular}
%\end{minipage}
\end{table*}
%%%%%%%%%%%%%%%%%%%%%%%%%%%%%%%%%%%%%%%%%%%%%%%%%%%%%%%%%%%%%%%%%%%%%%%%%%%

%%%%%%%%%%%%%%%%%%%%%%%%%%%%%%%%%%%%%%%%%%%%%%%%%%%%%%%%%%%%%%%%%%%%%%%%%%%
\begin{table*}
 \centering
 \begin{minipage}{150mm}
%  \caption{Stars selected for modelling.}
%  \caption{Target stars.}\label{t_st}
  \caption{Photometric and polarimetric data sources.}\label{t_st2}
  \begin{tabular}{lll} \hline
~~~Star &  {Photometry\footnote{%Source of the data:
%E81 is for \citet{Engels_1981},
M97 is for \citet{Mermilliod_1997}, 
W02 for  \citet{wegner_2002}, 
C03 for \citet{Cutri_2003},
M03 for \citet{Monet_2003},
V04 for \citet{valen_2004},
L05 for \citet{Larson_2005},
S05 for \citet{sofia_2005},
F07 for \citet{fm07}.}} & 
{Polarimetry\footnote{
C66 is for \citet{Coyne_1966},
S69 for \citet{Serkowski_1969},
%W80 for \citet{wilkin_1980}, 
%W82 for \citet{wilkin_1982},
A96 for \citet{ander_1996}.%, D07 for Das (2007).
}} \\
\hline
HD~24263 &  V04 (UV); M97 (UBV); M03 (RI); L05 (JHKL)                & A96 (UV \& vis--nearIR) \\
HD~62542 &  S05 (farUV); F07 (UV); M97 (BV); M03 (RI); C03 (JHK$_S$) & A96 (UV \& vis--nearIR)\\
HD~99264 &  F07 (UV); M97 (UBVRI); C03 (JHK$_S$)                     & A96 (UV \& vis); S69 (UBV) \\
%HD~147165&  F07 (UV); M97 (UBVRI); E81 (JHKLM)                       & this work (vis); C66 (vis--nearIR) \\
HD~147165&  {V04 (UV)}; W02 (UBVRIJHKLM)                               & this work (vis); C66 (vis--nearIR) \\
HD~147933&  F07 (UV); M97 (UBVRI); C03 (JHK$_S$)                     & A96 (UV--nearIR) \\ %; W86 (vis--nearIR?) \\
%HD~179406&  F07 (UV); M97 (UBV); M03 (RI); C03 (JHK$_S$)             & this work (vis) \\
HD~179406&  {V04 (UV)}; W02 (UBVJHKL); M03 (RI)                        & this work (vis) \\
HD~197770&  S05 (farUV); V04 (UV); M97 (UBV); M03 (RI); C03 (JHK$_S$)& A96 (UV--nearIR) \\ %; W80 (vis--nearIR?) \\
\hline
\end{tabular}
\end{minipage}
\end{table*}
%%%%%%%%%%%%%%%%%%%%%%%%%%%%%%%%%%%%%%%%%%%%%%%%%%%%%%%%%%%%%%%%%%%%%%%%%%%

\subsection{Observational data}\label{obs-d}

Basic information on the selected stars
is presented in Table~\ref{t_st1}.
%(the galactic coordinates $l, b$, 
 The distance $D$, colour excess $E(B-V)$, 
ratio of total extinction to the selective one
$R_{\rm V}$ and extinction $A_{\rm V}$
for all stars except HD 24263 and 197770
are from \citet{fm07}, with
the data for these two stars being %HD 24263 and 197770 
from \citet{valen_2004}. 
 The MK spectral types were taken from the 
catalogue of stellar spectral classifications 
(\citealt{Skiff_2009}).
 
The photometric data in the standard visual -- near infrared bands
were found in the general catalogue of photometric data
(\citealt{Mermilliod_1997}) and 
recent papers referred in the Simbad database (see Table~\ref{t_st2}).
 When the JHK data were unavailable, 
the JHK$_S$ values from the 2MASS catalogue (\citealt{Cutri_2003}) were used.
 When the RI data were absent, we applied with  caution the data
from the USNO B1 catalogue (\citealt{Monet_2003}).
 To derive the interstellar extinction in the line of sight
the intrinsic colours from \citet{Straizys_1992} were utilized.
  The UV extinction parameters were taken from
\citet{valen_2004} {and}  \citet{fm07}.  

{
Uncertainties of UV extinction are mainly a result of errors in $R_{\rm V}$ and $A_{\rm V}$ values. 
Note that typical relative errors of these quantites are 10--15\% (see, e.g., \citealt{valen_2004}).
An additional source of relatively large errors of extinction values in the visual and near-IR regions
is uncertainty of the intrinsic colors (cf., e.g., 
\citealt{Bessell_1988}; \citealt{Straizys_1992}; \citealt{Ducati_2001}). 
So, relative errors of extinction for individual objects are rather large (see, e.g., Fig.~2).
}

 Table~\ref{t_st2} also contains references to the papers with 
polarimetric data.
Additional polarimetric observations of HD~147165 and HD~179406 
were performed by one of the authors (HKD) in 2007 
at a 2m telescope of the Girawali Observatory (Pune, India).
 The IUCAA Faint Object Spectrograph and Camera (IFOSC) 
was used in the polarimetric mode.
 The polarization parameters were measured in 10 bands in the 
wavelength region about 0.37 -- 0.60\ $\mkm$ and 
a standard package was applied for data analysis. 

Note that 
for two stars (HD~62542 and HD~197770), 
far-UV extinction curves were derived from
IUE and FUSE observations by \citet{sofia_2005}.
For two stars (HD~147933 and HD~197770),
the {important} polarization data in the J, H and K bands are 
also available (\citealt{wilkin_1980}; \citealt{wilkin_1982}).

%================================================================
\subsection{Fitting procedure and  parameter variations}

%%%%%%%%%%%%%%%%%%%%%%%%%%%%%%%%%%%%%%%%%%%%
\begin{table*}
 \centering
 \begin{minipage}{170mm}
\caption{Modelling results for HD~147933.}\label{t933}
  \begin{tabular}{ccccccccccccccccccccccccccccc}
\hline
%\noalign{\smallskip}
\multicolumn{4}{c}{amorphous carbon}&\multicolumn{4}{c}{astrosil}&\multicolumn{1}{c}{graphite}
&\multicolumn{1}{c}{shape; } &&& \multicolumn{2}{c}{$\chi^2$} \\
%\cline{2-11} \cline{15-16}
  ${\cal K}_{\rm C}$  & $r_{V,\min}$\footnote{$r_{V,\min}$ and $r_{V,\max}$ are in $\mu$m}
  & $r_{V,\max}$ & $q$
 &${\cal K}_{\rm Si}$ & $r_{V,\min}$ & $r_{V,\max}$ & $q$
 &${\cal K}_{\rm gra}$& $a/b$               & $\delta_0 (\mu$m) & $\Omega$ (deg) & $A/A_{\rm V}$ & $P/A$  \\
\hline
0.24 & 0.03  & 0.15 & 1.5 & 0.73 & 0.08 & 0.25 & 1.5  & 0.03  & pro; 5 & 0.5 & 35 &1.04 &24.6 \\
0.54 & 0.10  & 0.20 & 2.0 & 0.40 & 0.01 & 0.25 & 1.3  & 0.06  & pro; 3 & 0.3 & 42 &0.19 &~8.2 \\
0.50 & 0.10  & 0.25 & 2.0 & 0.47 & 0.07 & 0.20 & 2.0  & 0.03  & obl; 3 & 0.3 & 33 &0.19 &~2.2 \\
\hline
%\vspace{5mm}
\end{tabular}
\end{minipage}
\end{table*}
%%%%%%%%%%%%%%%%%%%%%%%%%%%%%%%%%%%%%%%%%%%%

%%%%%%%%%%%%%%%%%%%%%%%%%%%%%%%%%%%%%%%%%%%%
\begin{figure*}

\includegraphics[angle=0, width=15cm]{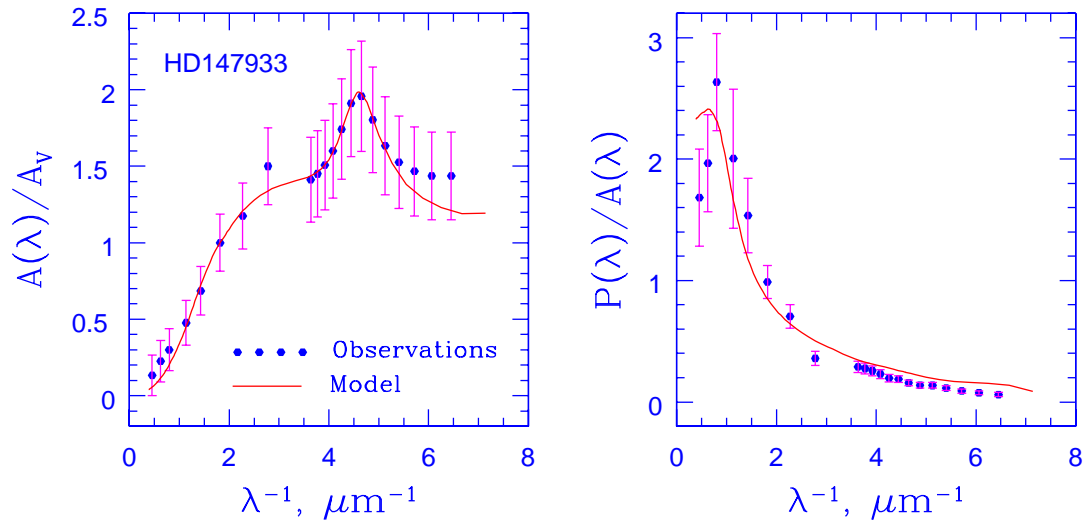}
 \caption{Comparison of the normalized extinction (left panel) and
the polarizing efficiency (right panel) observed for HD~147933 
with results of modelling.
 The solid curves show the case of prolate ($a/b=3$) spheroids having  
a power-law size distribution. 
 Other model parameters are given in Table~\ref{t933}.}
\label{147933p}
\end{figure*} 
%%%%%%%%%%%%%%%%%%%%%%%%%%%%%%%%%%%%%%%%%%%

A general problem of dust optics modelling is 
{\rm in obtaining the}
absolute values of the model parameters. 
 As it has been noted many times (e.g., \citealt{greenberg_1978}),
similar extinction occurs when a product of the typical
particle size $\langle r \rangle$ {and} the 
particle refractive index does not change, i.e.
$$
\langle r \rangle  \, |m-1| \approx {\rm const.}
$$
 The average Galactic extinction curve was successfully
reproduced using quite different mixtures of
particles with exponential (\citealt{hong_1980}),
power-law (\citealt{mathis_1977}) and more complicated
grain size distributions (\citealt{mathis_1996}; \citealt{weind_2001}).
 Effects of variations of the distribution parameters
(e.g., $r_{V, \min}$, $r_{V, \max}$, $q$ for a
power-law distribution) on extinction 
are well studied for spheres (see, for example, \citealt{vi93}).

Joint modelling of interstellar extinction and polarization adds
at least one parameter --- particle shape --- to the model.
 In the case of spheroids the particle shape is defined by the
spheroid's aspect ratio $a/b$. Besides this, we can consider
prolate (``cigars'') and oblate (``pancakes'') spheroids.
 If such particles are imperfectly oriented, the
alignment parameters (in our case, $\delta_0$ and $\Omega$) 
very weakly affect extinction. % at a undetectable rate. 
 Note that the normalized extinction $A_\lambda/A_V$ essentially
changes in the UV region when $a/b$ varies (see Fig.~3 in VD08).

In order to analyze importance of the model parameters
we reproduced the observational data available for HD~147933 
with different particles. 
 Some results are shown in
Figs.~\ref{otherm} and \ref{147933p} (see also Fig.~8 in VD08),
%Figs.~\ref{147933p} and \ref{147933o} (see also Fig.~8 in VD08),
the parameters of the fits are given in Table~\ref{t933}.
 Note that in VD08 the fitting was simply ``chi-by-eye" 
while in this paper we apply minimization of $\chi^2$.

%%%%%%%%%%%%%%%%%%%%%%%%%%%%%%%%%%%%%%%%%%%%
%\begin{figure*}
%
%\includegraphics[angle=0, width=15cm]{fig.eps}
%%\vspace{-70mm}
% \caption{The same, as Fig.~\ref{147933p}, but now for
%oblate spheroids with $a/b=3$.}
%\label{147933o}
%\end{figure*}
%%%%%%%%%%%%%%%%%%%%%%%%%%%%%%%%%%%%%%%%%%%%

The fitting approach used included two steps: 
i) finding of the size distribution
parameters and the relative column densities ${\cal K}_{i}$ 
from approximation of the extinction data 
using particles of given shape, and 
ii) a more accurate 
determination of the alignment parameters
$\delta_0$ and $\Omega$ from fitting of 
the wavelength dependence of polarizing
efficiency $P(\lambda)/A(\lambda)$.

 It is interesting that by varying the parameters
$r_{V, \min}$, $r_{V, \max}$ and $q$ and the relative density of carbon
(${\cal K}_{\rm C}$) and silicate (${\cal K}_{\rm Si}$) grains
one can reproduce the observational data with prolate as well as
oblate particles of different aspect ratios.
 Note that after fixing the particle type (prolate/oblate) and $a/b$ 
one can estimate only
{\rm the values of
${\cal K}_{\rm C}$, ${\cal K}_{\rm Si}$,
$r_{V, \min}$ and $q$ %which are obtained rather arbitrary.
%as $r_{V, \max}$ has little effect on the extinction and polarization curves.
as the extinction and polarization curves is little affected by
variation of  $r_{V, \max}$.}

The alignment degree for three models presented in Table~\ref{t933}
is found to be slightly larger than the standard interstellar value
($\delta_0 \approx 0.2\,\mu$m, \citealt{vosh_1989}). 
 This can be a result of enhanced magnetic fields 
in the interstellar cloud near $\rho$~Oph with which HD~147933 is related.
 The most exciting conclusion of our modelling is 
a close coincidence of the magnetic field directions ---
the values of $\Omega$ for various fits differ less than 10\degr.
 We suggest that this model parameter can be more or less reliably 
determined from such modelling.

Possibly, a better coincidence of the theory with 
the observations could be obtained if we consider
different alignment parameters for silicate and the
carbonaceous grains.
 Certainly, consideration of non-aligned carbonaceous grains cannot
be ruled out
({ \citealt{lg02, chiar_2005}; {\rm see, however, Fig.~\ref{otherm}}}).

%\clearpage
%{\rm -----------------------------------------------------}

\subsection{Fitting results and discussion}

The normalized extinction $A(\lambda)/A_{\rm V}$ and
the polarizing efficiency $P(\lambda)/A(\lambda)$
were calculated with the theory of Sect.~\ref{theor} 
and compared with the observational data available for 
the stars mentioned in Sect.~\ref{sel-o}.
 {\rm The results of this comparison are plotted in
Figs.~\ref{147933p}--\ref{f-res} 
(see also Fig.~9 in VD08 for HD~62542)
with the parameters of the fits being presented %in a tabular form
in Table~\ref{t_res}.
 We start with  comments on the individual objects and
then make general conclusions.}
%%%%%%%%%%%%%%%%%%%%%%%%%%%%%%%%%%%%%%%%%%%%
\begin{table*}
 \centering
% \begin{minipage}{140mm}
 \begin{minipage}{170mm}
 \caption{Modelling results.}\label{t_res}
 % \begin{tabular}{|l|ll|l|l|l|l|l|l|l|l|l|l|l|}
  \begin{tabular}{|ccccccccccccccccccccccccccc}
\hline
%\noalign{\smallskip}
&\multicolumn{4}{c}{amorphous carbon}&\multicolumn{4}{c}{astrosil}&\multicolumn{1}{c}{graphite}
&\multicolumn{1}{c}{shape;} \\
%\cline{2-11} \cline{15-16}
Star & ${\cal K}_{\rm C}$  & $r_{V,\min}$\footnote{$r_{V,\min}$ and $r_{V,\max}$ are in $\mu$m}
& $r_{V,\max}$ & $q$
     & ${\cal K}_{\rm Si}$ & $r_{V,\min}$ & $r_{V,\max}$ & $q$
     & ${\cal K}_{\rm gra}$& $a/b$               & $\delta_0 (\mu$m) & $\Omega$ (deg) \\
\hline
HD~24263  & 0.48 & 0.08  & 0.40 & 3.5 & 0.47 & 0.04 & 0.15 & 2.1  & 0.05  & pro; 2 & 0.3 & 60 \\
HD~62542  & 0.85 & 0.07  & 0.10 & 2.7 & 0.10 & 0.005& 0.25 & 2.7  & 0.05  & pro; 2 & 0.5 & 40 \\
HD~99264  & 0.44 & 0.09  & 0.30 & 6.0 & 0.50 & 0.04 & 0.18 & 2.0  & 0.06  & pro; 3 & 0.5 & 50 \\
HD~147165 & 0.49 & 0.06  & 0.30 & 2.2 & 0.44 & 0.04 & 0.20 & 2.2  & 0.07  & pro; 2 & 0.5 & 54 \\
HD~147933 & 0.50 & 0.10  & 0.25 & 2.0 & 0.47 & 0.07 & 0.20 & 2.0  & 0.03  & obl; 3 & 0.3 & 33 \\
HD~179406 & 0.45 & 0.10  & 0.40 & 6.0 & 0.47 & 0.04 & 0.18 & 2.5  & 0.08  & pro; 3 & 0.5 & 49 \\
HD~197770 & 0.43 & 0.05  & 0.20 & 2.8 & 0.46 & 0.01 & 0.17 & 1.3  & 0.11  & pro; 3 & 0.5 & 42 \\
\hline
%\vspace{5mm}
\end{tabular}
\end{minipage}
\end{table*}

%%%%%%%%%%%%%%%%%%%%%%%%%%%%%%%%%%%%%%%%%%%%
\begin{figure*}

\includegraphics[angle=0, width=15cm]{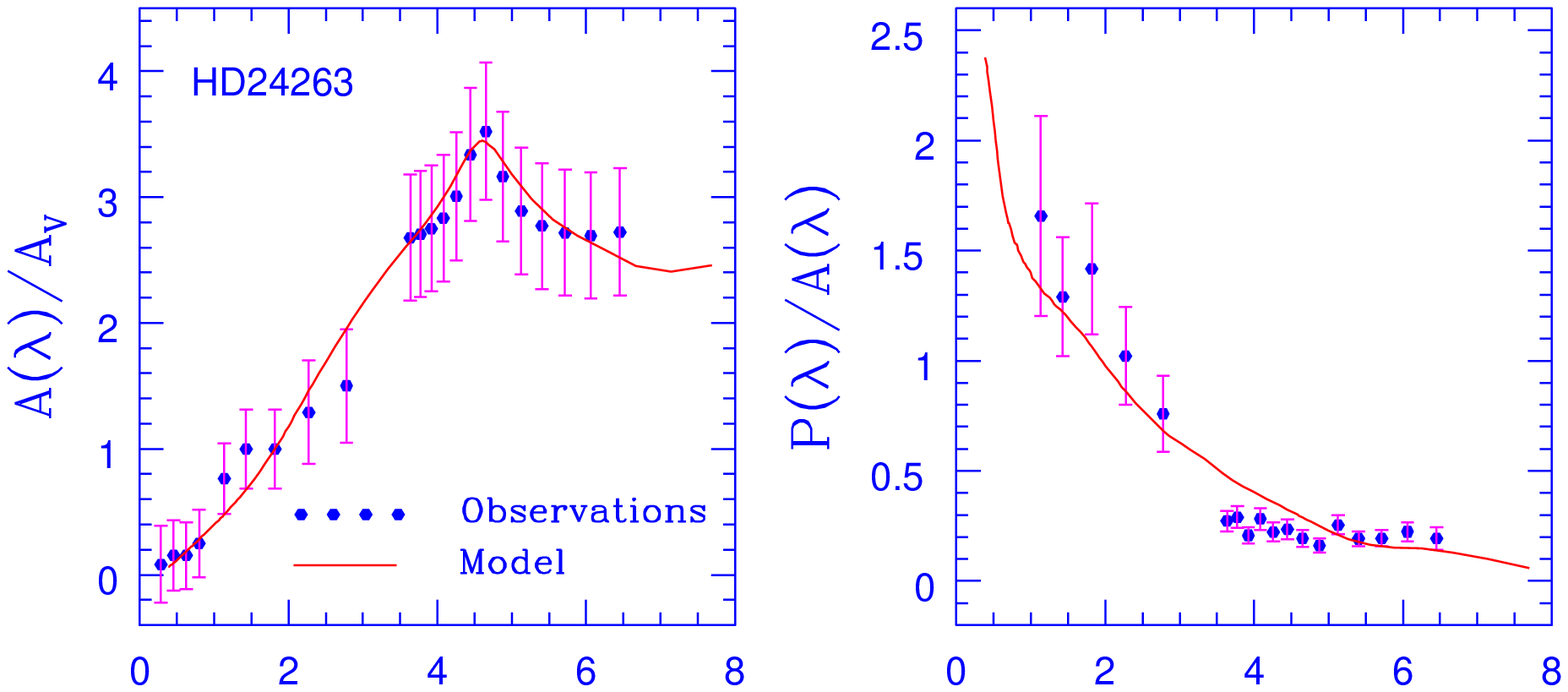}
   
\includegraphics[angle=0, width=15cm]{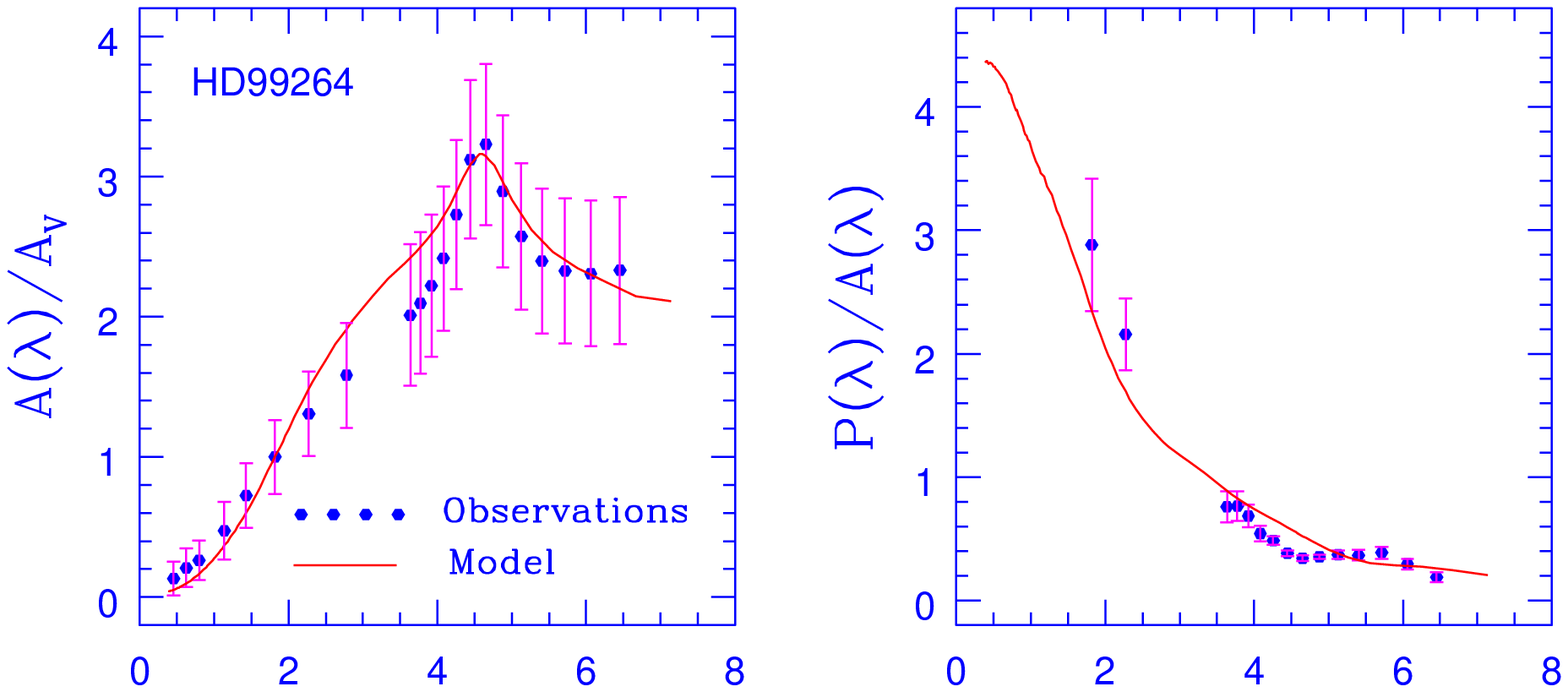}
   
\includegraphics[angle=0, width=15cm]{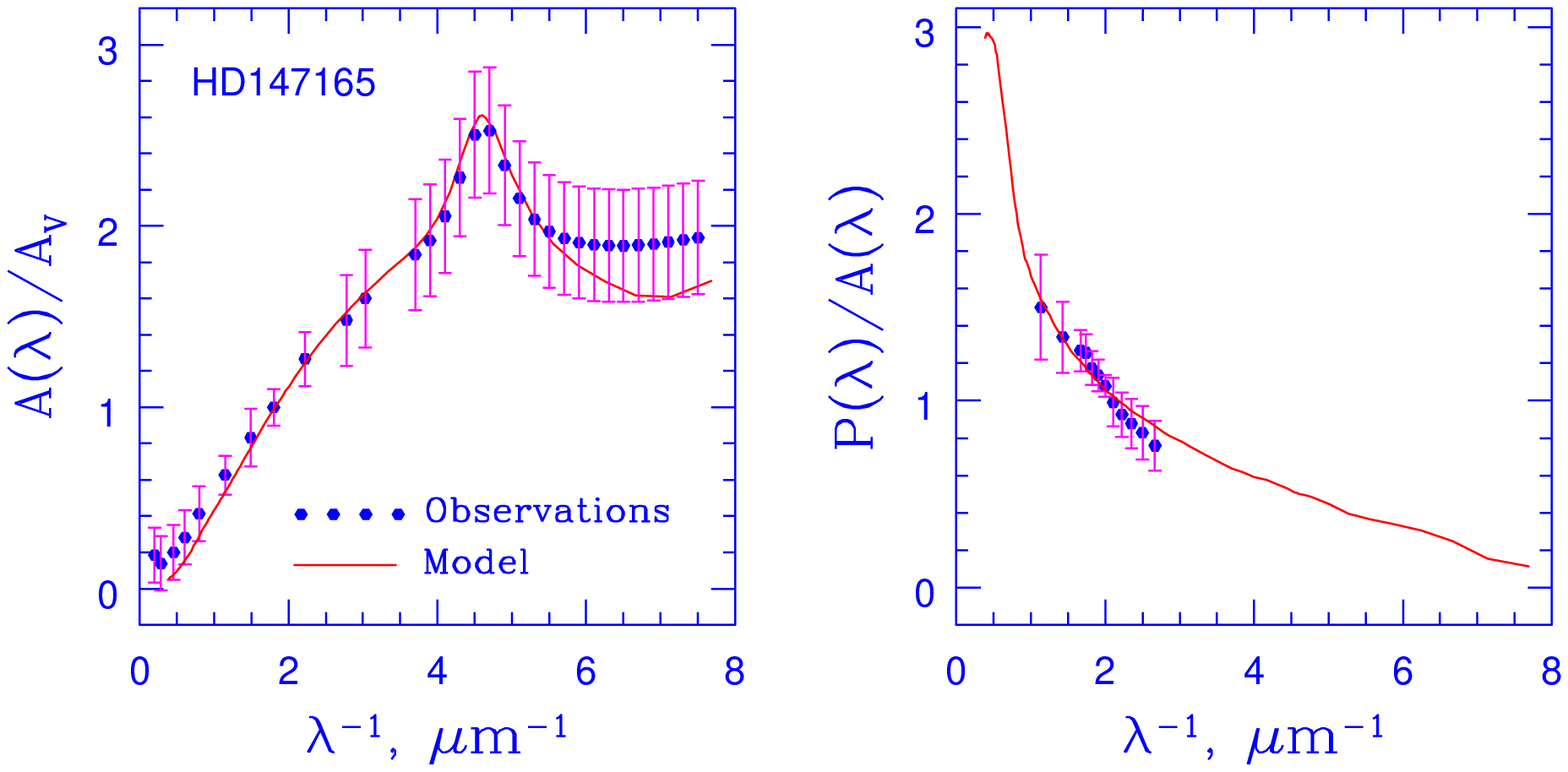}
\caption{Comparison of observed and modelled normalized extinction (left panel) and 
polarizing efficiency (right panel) for five stars.
 The solid curves show results of calculations.
The model parameters are given in Table~\ref{t_res}.
}\label{f-res}
\end{figure*}
%%%%%%%%%%%%%%%%%%%%%%%%%%%%%%%%%%%%%%%%%%%%
\setcounter{figure}{3}
%%%%%%%%%%%%%%%%%%%%%%%%%%%%%%%%%%%%%%%%%%%%
\begin{figure*}

\includegraphics[angle=0, width=15cm]{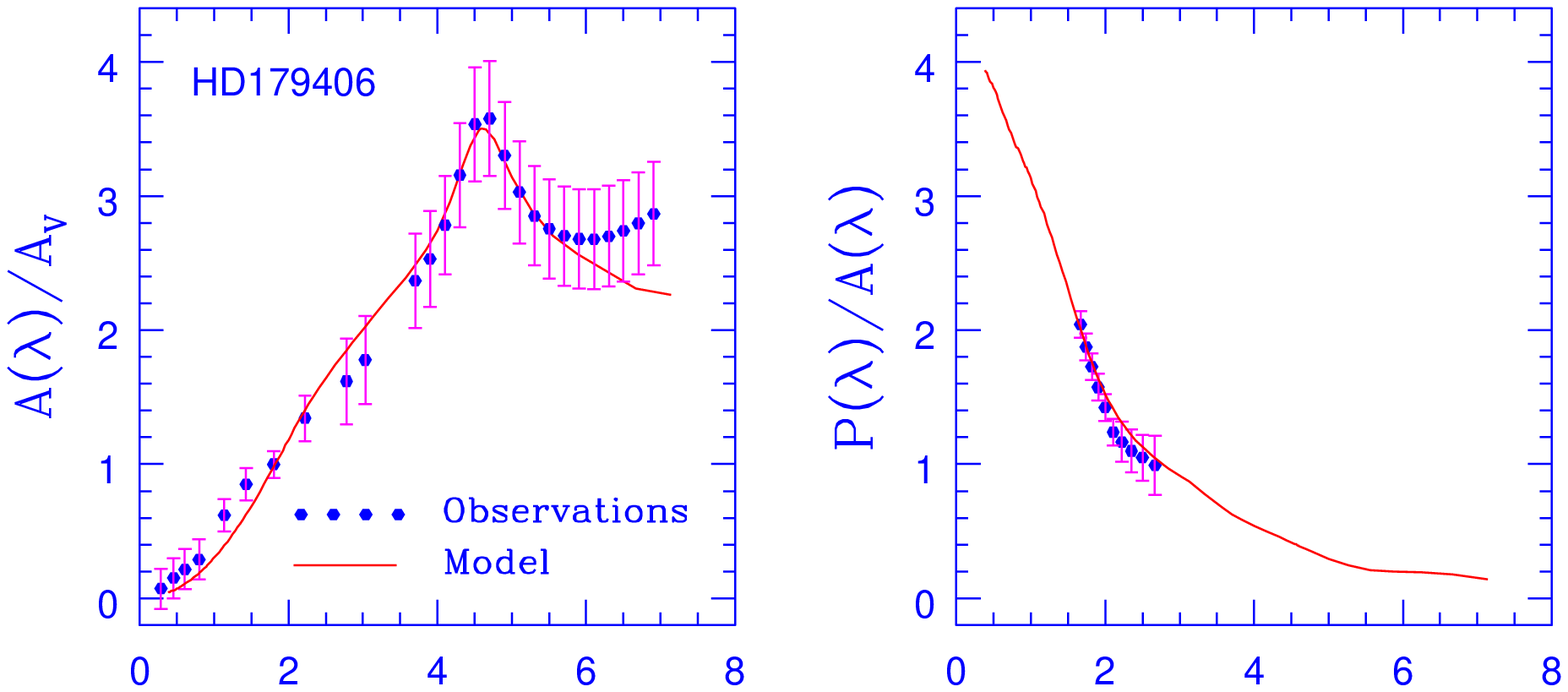}

\includegraphics[angle=0, width=15cm]{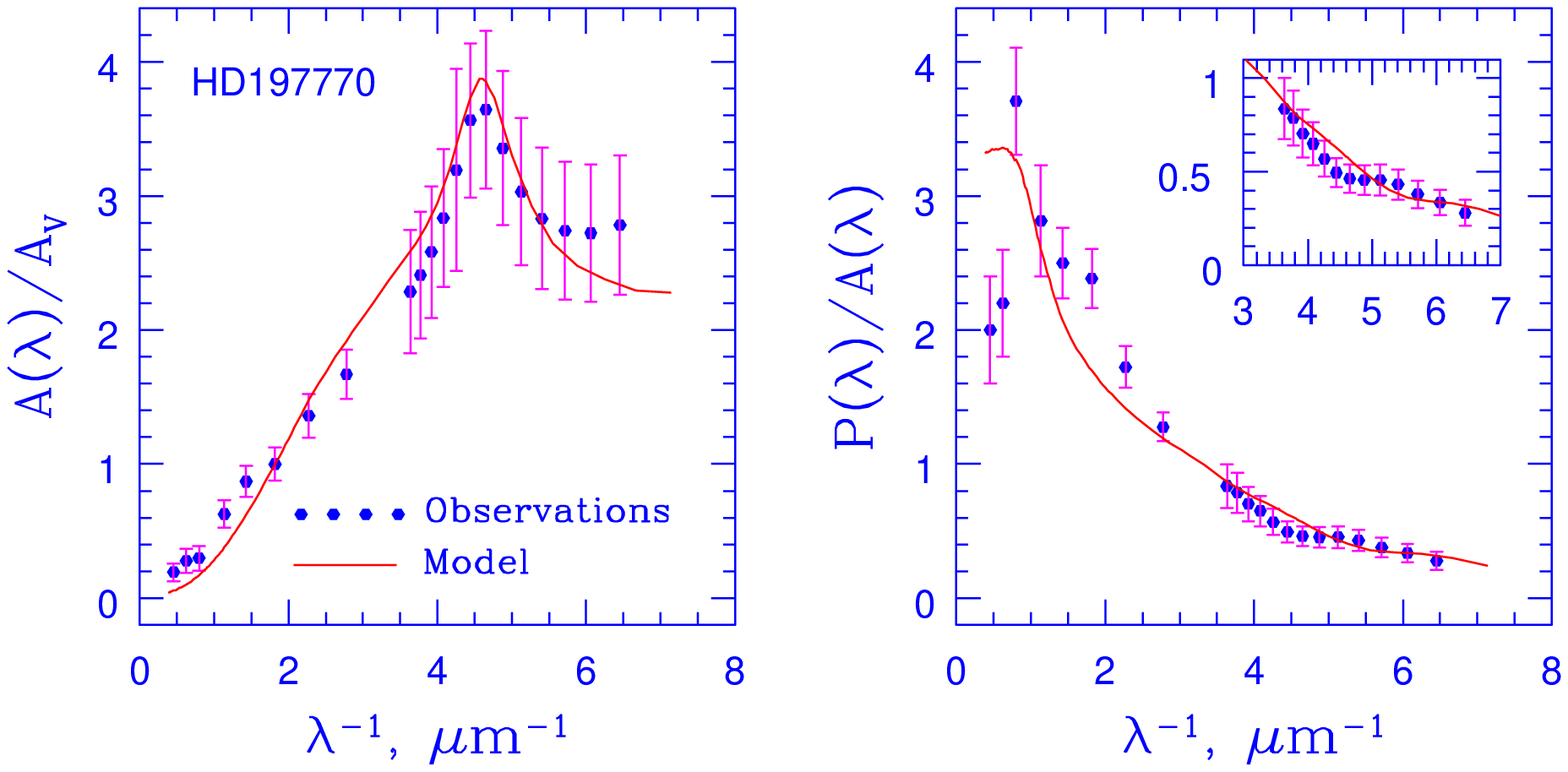}
 \caption{(continued).}
\end{figure*}

%%%%%%%%%%%%%%%%%%%%%%%%%%%%%%%%%%%%%%%%%%%%
%\subsubsection{Individual objects}
%%%%%%%%%%%%%%%%%%%%%%%%%%%%%%%%%%%%%%%%%%%%

%\smallskip
\noindent{\it HD~62542}. 
The UV extinction curve has
a very broad and weak 2175 \AA\, bump and a steep far-UV rise
(\citealt{snow06}). Our model suggests that the major part
of grains are small carbon particles with a very narrow size
distribution.

%\smallskip
\noindent{\it HD~99264}.
 The extinction curve for it is characterized
by $R_{\rm V}$ of about 3 {\rm \citep{fm07}},
the UV extinction curve is a bit ``shallow'' but not anomalous
\citep{Carnochan_1986}.
 The polarization in an UV region and the UBV bands
is well described by Serkowski curve
with $\lambda_{\max} = 0.55 \mkm$
(\citealt{ander_1996, Martin_1999}).
 So, we have the case close to the average Galactic curves,
% but certainly there are not enough data for good fitting.
{\rm  but certainly there are not enough
polarization data for sound comparison.}

%\smallskip
\noindent{\it HD~147165} and {\it HD~179406}.
Extinction in these lines of sight along with
{\rm  element abundances}
were interpreted by \citet{zubko_1996, zubko_1998}
using rather wide size distributions of grains. 
 These authors applied spherical particles of different materials as well as
inhomogeneous particles, which makes impossible a
{\rm direct} comparison of
their results with those obtained in our work. 

%\smallskip
\noindent{\it HD~197770}.
For this star the polarizing efficiency is maximal: it reaches $P/A=3.7$ at
the J band ($\lambda= 1.25\,\mu$m).
The polarization is enhanced in the region of the UV bump
(super-Serkowski behaviour) but strong UV bump in extinction {results}
in no structure for $P/A$ curve
{ (see insert in Fig.~\ref{f-res}).}
Our model shows the largest fraction of
graphite in the grain mixture for this star.

%\afterpage{\clearpage}

%\subsubsection{General discussion}
%\smallskip
Thus, we have fitted the observational data using
prolate or oblate spheroids with $a/b=2$ or 3 and 
 intermediate values of the alignment parameter $\delta_0 =0.3 - 0.5\,\mu$m.
 Contributions of carbonaceous and silicate grains to extinction
and polarization are nearly equal excluding the case of HD~62542 
for which extinction is very peculiar.
 A contribution of small graphite particles to extinction 
usually is less than 10\%.
 The parameters of the size distributions for silicate and amorphous
carbon grains differ from the standard MRN mixture
(see Sect.~\ref{s-z}). 
 As usual, very small particles with
$r_{V, \min}\la 0.01 \,\mu$m are absent in the grain ensembles.
 Besides this, in many cases the slope of the $n(r_{V})$ dependence 
is less steep than that for the MRN mixture (i.e. $q<3.5$).

The inclination of the magnetic field direction relative to the line
of sight determined for all target stars lies in the interval 
$\Omega \approx$ 30  -- 60\degr. These values of $\Omega$
are derived rather accurately because of a strong dependence of
$P/A$ on $\Omega$ (see Figs.~4 and 6 in VD08). The magnetic
fields seem to be connected with local dust clouds in the directions
to the selected stars. 
 At the same time, the obtained values of $\Omega$
do not contradict the general pattern of the galactic magnetic
field in the solar neighbourhood as found from pulsar measurements
%(\citealt{val04, val08}; \citealt{han07}).
(\citealt{val04, val08}).

{
 It should be noted that one usually assumed that small grains were
badly oriented as the interstellar polarization degree decreases
with a growing $\lambda^{-1}$ in the UV region (\citealt{whitt_2003}). 
 However, for all the objects we have fitted
the observational data by using the simple size and 
orientation distributions 
with small particles being rather well aligned.
 To see that one should consider Eqs.~(\ref{eq5})--(\ref{eq6}) 
and keep in mind that $\delta_0$ is about 0.4 and
the temperature ratio %$T_{\rm d}/T_{\rm g}$ 
about 0.1. 
 For $r_{\rm V} \sim 0.06 \mkm$, this gives $\xi \sim 0.5$
when $\xi = 0$ corresponds to the perfect alignment 
while $\xi = 1$ means the random orientation.   
}

%\clearpage

\section{Conclusions}

{
We have extended the spheroidal model of interstellar dust by
considering imperfectly dynamically aligned amorphous carbon and
silicate prolate and oblate spheroids with a power-law size distribution
and a small fraction of graphite spheres involved to explain the 2175\AA\, bump.
%the $\lambda$2200 bump.
 Our model being as simple as possible was applied to fit the
the normalized extinction $A(\lambda)/A_{\rm V}$ and the 
polarizing efficiency $P(\lambda)/A(\lambda)$ observed in
the near IR to far UV region.

 In contrast to earlier works 
we exactly calculated the optical properties of spheroids, 
for the first time utilized their imperfect Davis--Greenstein alignment and  
applied the spheroidal model to fit the data obtained for individual stars
instead of the commonly used average Galactic curves. 

A comparison of our model with those recently suggested by other authors
has demonstrated that the extinction data are well fitted by all the models
while only our { model} allows one to satisfactorily fit the polarization
data { with simple size and orientation} distributions.  
 {  Contrary to other works we find that the data can be well fitted with
aligned carbonaceous and small silicate non-spherical grains.}

Investigation of our model has shown that the only parameter that can be
determined more or less accurately from observational data fitting
is the direction of magnetic field relative to the
line of sight. % characterized by the angle $\Omega$. 
 As the projection of the magnetic
field on the sky plane is given by the positional angle of
polarization $\vartheta$,
multi-wavelength extinction and polarization
observations can be used for
study of the spatial structure of interstellar magnetic fields.

Variations of the interstellar extinction and polarization curves
for the considered stars can be interpreted by differences 
in the relative fractions of carbon and silicate grains 
and their size distributions in the clouds crossed by the lines of sight
with the alignment parameter having intermediate values 
$\delta_0 = 0.3 - 0.5 \mkm$ (for the particle aspect ratio $a/b = 2 - 3$).
%It is found that the observations can be fitted with models
%having essentially different fractions of carbon and silicate grains and
%size distribution parameters. 
{ We assume that} ambiguous results could be ruled out by considering dust phase abundances.
}

%\end{itemize}

%\acknowledgments
\section*{Acknowledgments}
%We are thankful to Vladimir Il'in for careful reading of manuscript.
One of the authors (HKD) is deeply indebted to the IUCAA's instrumentation 
and observatory team for their support and cooperation extended 
during the course of this work.
The work was partly supported by the grants
RFBR 07-02-00831, NTP 2.1.1/665 and NSh 1318.2008.2.
%of the Russian Federation.

%@@@@@@@@@@@@@@@@@@@@@@@@@@@@@@@@@@@@@@@@@@@@@@@@@@@@@@@@@@@@@@@@@@@@@@@@@@@@@@@@@@@

%%%%%%%%%%%%%%%%%%%%%%%%%%%%%%%%%%%%%%%%%%%%%%%%%%%%%%%%%%%%%%
% ALL ADDED BY V.I. BIBITEMS BEGIN WITH THE THIRD POSITION!!!
%%%%%%%%%%%%%%%%%%%%%%%%%%%%%%%%%%%%%%%%%%%%%%%%%%%%%%%%%%%%%% 

%********************************************************************
%%%%%%%%%%%%%%%%%%%%%%%%%%%%%%%%%%%%%%%%%%%%
\label{lastpage}
\end{document}